\documentclass[aps,amsmath,amssymb,prl,twocolumn,10pt,showpacs]{revtex4-1}
\usepackage{graphicx}
\usepackage{bm,txfonts}
\usepackage{mathrsfs}
\usepackage{comment}

\newcommand{\al}{\alpha}
\newcommand{\be}{\beta}

\newcommand{\de}{\delta}
\newcommand{\e}{\epsilon}
\newcommand{\z}{\zeta}

\newcommand{\thi}{\theta}

\newcommand{\la}{\lambda}

\newcommand{\p}{\pi}

\newcommand{\s}{\sigma}

\newcommand{\f}{\phi}
\newcommand{\x}{\chi}
\newcommand{\w}{\omega}
\newcommand{\W}{\Omega}
\newcommand{\De}{\Delta}
\newcommand{\G}{\Gamma}

\newcommand{\pd}{\partial}

\renewcommand{\square}[1]{\left[ #1 \right]}

\newcommand{\abs}[1]{\left| #1 \right|}

\newcommand{\ang}[1]{\left\langle #1 \right\rangle}

\newcommand{\beq}{\begin{equation}}
\newcommand{\eeq}{\end{equation}}
\newcommand{\Beq}{\begin{eqnarray}}
\newcommand{\Eeq}{\end{eqnarray}}
\newcommand{\bml}{\begin{multline}}

\newcommand{\bea}{\begin{align}}
\newcommand{\ena}{\end{align}}
\newcommand{\bsp}{\begin{split}}
\newcommand{\esp}{\end{split}}
\newcommand{\down}{\downarrow}
\newcommand{\up}{\uparrow}

\newcommand{\br}{{\boldsymbol r}}

\newcommand{\bS}{{\boldsymbol{S}}}

\newcommand{\bmus}{{\boldsymbol \mu}_s}
\newcommand{\tbmus}{\tilde{{\boldsymbol \mu}}_s}
\newcommand{\ey}{{\boldsymbol e_y}}
\newcommand{\ez}{{\boldsymbol e_z}}
\newcommand{\bi}{{\boldsymbol i}}
\newcommand{\bj}{{\boldsymbol j}}
\newcommand{\bJ}{{\boldsymbol J}}

\newcommand{\bk}{{\boldsymbol k}}

\newcommand{\bq}{{\boldsymbol q}}

\newcommand{\bQ}{{\boldsymbol Q}}

\newcommand{\sG}{\mathcal{G}}
\newcommand{\hH}{\hat{H}}
\newcommand{\hS}{\hat{S}}

\newcommand{\hphi}{\hat{\varphi}}
\newcommand{\bx}{\boldsymbol{x}}
\newcommand{\bn}{\boldsymbol{n}}
\newcommand{\by}{\boldsymbol{y}}
\newcommand{\hpsi}{\hat{\psi}}
\newcommand{\hV}{\hat{V}}
\newcommand{\hn}{\hat{n}}
\newcommand{\ve}{\varepsilon}
\newcommand{\bs}{{\boldsymbol{s}}}

\newcommand{\hJ}{\hat{J}}
\newcommand{\ha}{\hat{a}}

\newcommand{\tS}{\tilde{S}}

\begin{document}
\title{Superfluid Spin Transport through Easy-Plane Ferromagnetic Insulators}
\author{So Takei}
\affiliation{Department of Physics and Astronomy, University of California, Los Angeles, California 90095, USA}
\author{Yaroslav Tserkovnyak}
\affiliation{Department of Physics and Astronomy, University of California, Los Angeles, California 90095, USA}
\date{\today}

\begin{abstract}
Superfluid spin transport | dissipationless transport of spin | is theoretically studied in a ferromagnetic insulator with
easy-plane anisotropy. We consider an open geometry where spin current is injected into the ferromagnet from one side by a metallic reservoir
with a nonequilibrium spin accumulation, and ejected into another metallic reservoir located downstream. 
Spin transport is studied using a combination of magnetoelectric circuit theory, Landau-Lifshitz-Gilbert phenomenology, and microscopic linear-response theory. We discuss how spin superfluidity can be probed in a magnetically-mediated negative electron-drag experiment.
\end{abstract}
\maketitle

{\em Introduction}.|An important goal in the field of spintronics is to understand how spin, a quantum-mechanical unit of magnetism, can be exploited
for information transport, data storage and processing. 
While conventional spintronics \cite{zuticRMP04}, relying solely on conduction electrons in metals and semiconductors as carriers of spins,
still faces difficulties associated with fast spin relaxation and significant Joule heating, 
a promising alternative that combines conventional spintronics with coherent spin-wave dynamics in magnetic insulators has recently emerged \cite{khitunIEEEM08}.
Magnetic insulators 
can also transport spin information via magnons \cite{kajiwaraNAT10,*uchidaNATM10}, the quantum of spin waves that also carries a unit of angular momentum.
This emerging field of magnon spintronics may alleviate the obstacles present within the conventional schemes. The possibility to investigate spin transport in magnetic insulators also opens a new venue for their experimental probes.

Integrating magnetic insulators into spintronic devices raises interesting possibilities that stem from the bosonic nature of
the spin-carrying magnons. These magnons can form a 
Bose-Einstein condensate, which 
has been observed in some magnetic insulators including TlCuCl$_3$~\cite{nikuniPRL00,*oosawaJPCM99}, Cs$_2$CuCl$_4$~\cite{raduPRL05}, and
Y$_3$Fe$_5$O$_{12}$ (YIG) films \cite{demokritovNAT06}. A closely related phenomenon is superfluidity, which is another general property of bosonic quantum matter
at low temperatures. In magnetic systems, this raises the possibility of spin superfluidity, i.e., a dissipationless macroscopic transport of spin \cite{halperinPR69}.
In the past, the concept was used to explain unusually fast spin relaxation in $^3$He-A \cite{corrucciniPRL75,*vuorioJPC74,*vuorioJPC76,*soninJETPL79} and invoked to interpret the coherence of a nonuniformly precessing state of $^3$He-B \cite{borovikJETPL84,*fominJETPL84}. Spin superfluidity has also been
studied in Bose-condensed excitonic fluids \cite{soninSSC78}. While the absence of strict conservation laws for spin rules out faithful analogy
to conventional mass superfluidity \cite{guseinovJETPL73}, it was demonstrated that the analogy can still be useful if the violation of conservation law is weak \cite{soninSSC78}.
Generation of dissipationless spin current has received attention in the past in metallic systems with noncollinear 
magnetic order~\cite{konigPRL01}, $p$-doped semiconductors~\cite{murakamiSCI03}, and two-dimensional electron systems with Rashba 
spin-orbit coupling~\cite{sinovaPRL04}. 

\begin{figure}[t]
\centering
\includegraphics*[scale=0.38]{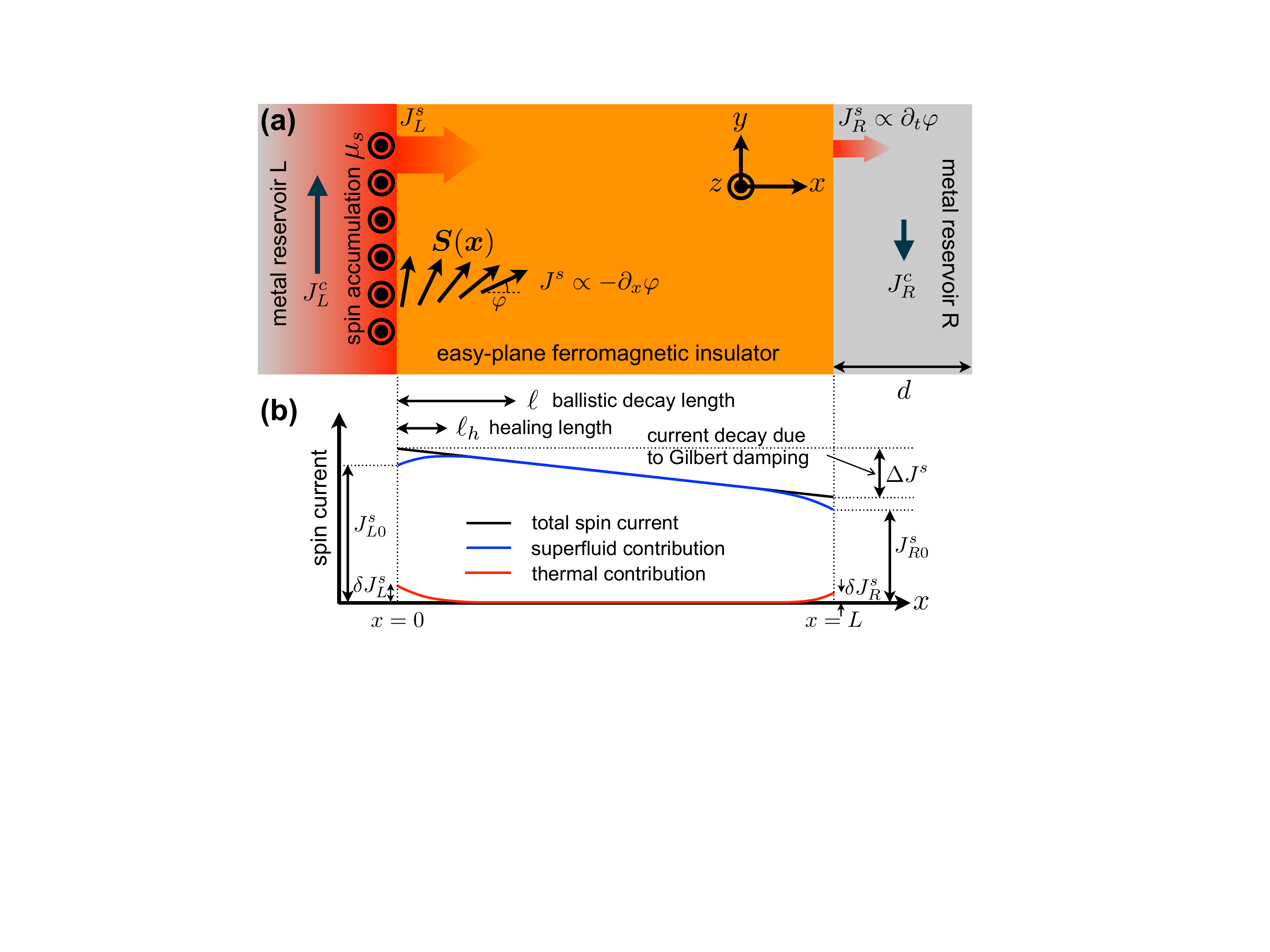}
\caption{(a) Schematic of the hybrid structure for realizing spin superfluidity. (b) A schematic plot showing the spatial distribution of the condensate and thermal contributions to the spin currents in the presence of Gilbert damping. See text for a detailed discussion.
}
\label{setup}
\end{figure}

In this Letter, we theoretically investigate how superfluid spin transport can be realized and detected in magnetic-insulator based hybrid structures. The notion of superfluid spin transport here is closely related to Ref.~\cite{konigPRL01}. In this work, we focus on the pertinent spin-transfer physics at the ferromagnet$|$metal interfaces (including thermally-activated spin currents), which is related to well-established and independently measurable quantities such as the spin-mixing conductance and the spin Hall angle. We identify the importance of global magnetic precession and the associated relaxation of spin superfluid by Gilbert damping.
Specifically, we consider a ferromagnetic insulator with easy-plane anisotropy 
attached on the two sides by metallic reservoirs that act as the source
and the drain for spin current (see Fig.~\ref{setup}). 
In an open geometry, superfluid spin transport is achieved by maintaining a spiral magnetic texture in the ferromagnet, 
along with a self-consistent magnetic precession within the easy plane, through a steady injection of angular momentum at the source and its 
depletion by spin pumping \cite{tserkovPRL02sp,*tserkovRMP05} at the drain.
The spin injection at the source and its ejection at the drain have two contributions: coherent spin-transfer torque \cite{slonczewskiJMMM96} and thermally-activated spin current mediated by magnons. We establish the length scales involved in the conversion of the thermal contribution into a superflow, with its eventual relaxation due to Gilbert damping. The resultant spin current can be probed in a magnetically-mediated negative electron-drag experiment, similar to the proposal in Ref.~\cite{zhangPRL12}, facilitated by the spin Hall effect at the ferromagnet$|$normal-metal contacts. 

{\em Superfluid spin transport.}|Before pursuing a more rigorous microscopic calculation, we first outline the essential semiclassical aspects of superfluid spin transport (see Fig.~\ref{setup}). We start at zero temperature, where spin current in the ferromagnet cannot be transported by magnons. A ferromagnet 
of length $L$ (occupying $0<x< L$) is sandwiched between two metallic reservoirs occupying $-\infty<x< 0$ and $L<x<\infty$. 
We assume full translational symmetry along the interface ($yz$) plane, axial symmetry about the $z$ axis, and take the easy plane in the $xy$ plane. 
The energy for the ferromagnet can be written as $H_F=\int d^3\bx[A(\nabla\bn(\bx))^2+Kn_z(\bx)^2]/2$,
where $A$ and $K$ parameterize the exchange stiffness and anisotropy, respectively, and $\bn(\bx)$ is the unit vector along the local spin density $\bs(\bx)$.
We parameterize $\bn$ by the azimuthal angle $\varphi$ and its $z$ projection $\bn=(\sqrt{1-n_z^2}\cos\varphi,\sqrt{1-n_z^2}\sin\varphi,n_z)$,
and describe its classical dynamics using the Landau-Lifshitz-Gilbert (LLG) equation $(1+\alpha\bn\times)\dot\bn=-\bn\times\pd_\bs H_F$, where $\al$ is a dimensionless damping constant that accounts for magnetic losses in the ferromagnet bulk. In the strong-anisotropy
and long-wavelength limit (i.e. $\lambda\gg\sqrt{A/K}$), the LLG equation can be expanded to lowest order in $n_z$ and gradients of $\varphi$:
\beq
\label{EOMdamp}
\dot\varphi=Kn_z/s+\al\dot n_z,\quad \dot n_z=A\nabla^2\varphi/s-\al\dot\varphi,
\eeq
where $s\equiv|\bs(\bx)|$ is assumed to be fixed at its saturation value. For $\al=0$, Eqs. (\ref{EOMdamp}) are a magnetic analog of the Josephson relations for superfluidity. The first term on the right hand side of the second equation defines the supercurrent density (for the $z$ projection of spin) as $\bJ^s(\bx)=-A\nabla\varphi(\bx)$, and the spin waves
can be shown to have a soundlike linear spectrum as in a superfluid with the speed $v=\sqrt{AK}/s$. From Eqs.~(\ref{EOMdamp}), we identify $\varphi$ and $sn_z$
as canonically-conjugate variables, with the long-wavelength Hamiltonian given by $H_F\approx\int d^3\bx\left[A(\nabla\varphi)^2+Kn_z^2\right]/2$.

Perturbing a monodomain ferromagnet by a nonequilibrium $z$-axis spin accumulation in the left reservoir, the magnet's spins cant out of the plane and acquire a nonzero $n_z$, which in turn triggers a 
precession of the magnetic order about the $z$ axis by virtue of Eqs.~(\ref{EOMdamp}). A steady-state solution for $n_z$ and $\varphi$ can be written as $n_z(\bx,t)\equiv\mbox{const}=n_z$ and $\varphi(\bx,t)=\varphi(x)+\W t$, where $\W=Kn_z/s$ is the precession frequency and $\varphi''=(\alpha s/A)\Omega$. According to the translational symmetry along the interfaces, we are considering solutions that are
independent of $(y,z)$. The magnetization canting, $n_z$, and the condensate spin current density flowing through the ferromagnet in the $x$ direction, $J^s$, must be found according to the appropriate boundary conditions at $x=0,L$. To that end, we employ the magnetoelectric circuit theory \cite{tserkovPRL02sp}, as follows.


The spin-current density injected into the ferromagnet from the left reservoir is given by $\bJ^s_L=(\Im g_L^{\up\down}+\Re g_L^{\up\down}\bn\times)(\tbmus\times\bn)/4\p$,
where $\Re g_r^{\up\down}$ and $\Im g_r^{\up\down}$ are respectively the real and imaginary parts of the spin-mixing conductance, $g_r^{\up\down}\equiv\Re g_r^{\up\down}+i\Im g_r^{\up\down}$,
associated with the ferromagnet$|$reservoir-$r$ interface.
$\tbmus$ has two contributions:
$\tbmus\equiv\bmus-\hbar\bn\times\dot\bn$, corresponding to spin-transfer torque and spin pumping, respectively.
Here, $\bmus=\mu_s\ez$ is the nonequilibrium spin accumulation in the left reservoir.
We thus have, in our linearized theory:
$\bn\times\dot\bn\approx\W\ez$ and $J^s_L=g_L^{\up\down}(\mu_s-\hbar\W)/4\p$ for the $z$ component of spin current. 
A similar analysis at the right interface gives the spin current injected into the right reservoir: $J^s_R=g_R^{\up\down}\hbar\W/4\p$. 
With finite damping, the amount of spin supercurrent dissipated in the ferromagnet
of length $L$ is given by $\Delta J^s\equiv J^s_L-J^s_R=\al s\W L$. 
Imposing the continuity of spin current at the boundaries, according to the circuit theory, we then obtain
\beq
\label{mainresult}
\W=\frac{\mu_s}{\hbar}\frac{g_L^{\up\down}}{g_L^{\up\down}+g_R^{\up\down}+g_\alpha}\,,\quad
J^s_R=\frac{\mu_s}{4\p}\frac{g_L^{\up\down}g_R^{\up\down}}{g_L^{\up\down}+g_R^{\up\down}+g_\alpha}\,,
\eeq
where $g_\alpha\equiv4\p\al sL/\hbar$. This is a central result of this work. Note that the supercurrent decays algebraically as a function of the ferromagnet's length 
$L$ in the presence of Gilbert damping. For spin transport mediated solely by magnons~\cite{zhangPRL12}, 
spin current is expected to decay exponentially over the magnon diffusion length $\la_{\rm sd}\sim v\sqrt{\tau\tau^*}$, $\tau$ ($\tau^*$) being the decay (scattering) mean free times. The detection of appreciable spin current for $L\gg\la_{\rm sd}$ should be an evidence of spin superfluidity.

{\em Microscopic theory}.|In order to account for finite-temperature corrections to the above results, we proceed to develop a linear-response theory for a concrete microscopic model. To that end, consider a ferromagnet with spins arranged on a cubic lattice. With the $xy$ easy plane, its energy
is $\hat{H}_F=-(J/2)\sum_{\ang{\bi\bj}}\bS_\bi\cdot\bS_\bj+(D/2)\sum_\bi S^2_{z\bi}$,
where $J>0$ is the exchange integral, $D>0$ is the anisotropy energy, and $\bS_\bi$ is the local spin in units of $\hbar$. 
Sites are labeled by $\bi,\bj$, and nearest-neighbor sites are denoted by $\ang{\bi\bj}$. 
The low-energy behavior of the system is described by replacing the spin $\bS_\bi$ on lattice site $\bi$ with
a continuum spin density, $\bS(\bx)\approx\bS_\bi/a^3$, that varies slowly in space. Owing to the axial symmetry about the $z$ axis, it is useful to 
parametrize the spin density using two slowly-varying fields, its azimuthal angle $\varphi(\bx)$ and $z$ component 
$S_z(\bx)$.
Retaining terms up to quadratic order in small quantities, the long-wavelength (quantum) Hamiltonian, in the case of a strong easy-plane anisotropy, 
becomes
\beq
\label{HFI}
\hH_F\approx\int d^3\bx\left[A\left(\nabla\hphi\right)^2+K\hn_z^2\right]/2\,,
\eeq
where $A=JS^2/a$, $K=DS^2/a^3$, and $\hn_z(\bx)=a^3\hS_z(\bx)/S$. 
The fields $\hphi(\bx)$ and $\hS_z(\bx)$ are canonically-conjugate variables obeying $[\hphi(\bx),\hS_z(\by)]=i\de(\bx-\by)$.
We have dropped terms in Eq.~(\ref{HFI}) that are higher order in $J (a/\lambda_T)^2/D\ll 1$, where $\lambda_T$ is the thermal magnon wavelength 
at temperature $T$. Using the long-wavelength magnon velocity $v=\sqrt{JD}aS/\hbar$ and the thermal
wavelength $\lambda_T\sim \hbar v/k_BT$, the above inequality gives a condition on the relevant temperature regime:
$T\ll DS/k_B\equiv T_D$. We generalize our results to the opposite, high-temperature regime, $T_D\ll T$ (where magnons become circular), at a later point.

The metallic reservoirs on the left ($r=L$) and right ($r=R$) are modeled as free electron gases with dispersion $\e_{\bk k_x r}=\hbar^2(|\bk|^2+k_x^2)/2m_r$,  
where $m_r$ denotes the effective electron masses for the two reservoirs, and $\bk$ ($k_x$) labels the wave number parallel (normal) to the interface plane. The nonequilibrium spin accumulation is induced in the left reservoir, where the chemical potentials of the two spin species are separated by $\mu_s$. The spin-dependent distribution function for the left reservoir is thus given by $n_{\rm FD}(\ve-\s\mu_s/2)$, where $n_{\rm FD}(x)=[e^{\be (x-\mu)}+1]^{-1}$, with
$\be=(k_BT)^{-1}$ and chemical potential $\mu$, and $\s=+(-)$ corresponds to the up-spin (down-spin) electrons.
Both spin species in the right reservoir obey $n_{\rm FD}(\ve)$.
We take the spin quantization axis for the spin accumulation along the $z$ axis. (For the case of in-plane spin accumulation, see Supplementary Material.)


We suppose that the ferromagnet magnetization and electron spin density at each interface couple via an $sd$-type 
exchange interaction. For strong anisotropy, the interaction Hamiltonian up to $\mathcal{O}(\hS_z/S)$ can be written as $\hV=\sum_r\hV_r$, with
\beq
\label{exchV}
\hV_r=\eta_r\int d^2\br\square{\frac{\tS e^{-i\hphi(x_r)}}{2}\hat s^+_r(x_r)+{\rm H.c.}
+\hS_z(x_r)\hat s^z_r(x_r)}\ ,
\eeq
where $\eta_r$ is the exchange coupling, $\tS=S/a^3$ and $x_{L,R}=0,L$. The reservoir spin densities 
are defined by $\hat s^i_{r}(\bx)=\hpsi^\dag_{\s r}(\bx)\tau^i_{\s\s'}\hpsi_{\s' r}(\bx)/2$,
where $\hpsi_{\s r}(\bx)$ is the annihilation operator for a spin-$\s$ electron in reservoir $r$ at position $\bx$, $\tau^i$ are the Pauli matrices and $\hat s^\pm_r=\hat s^x_r\pm i\hat s^y_r$. Here, we implicitly assume the dependence of the operators on $\br$.

The spin accumulation in the left reservoir leads to an injection of spin current in the form of 
a superfluid and a thermally-activated spin current mediated by magnons. 
Within the healing length, $\ell_h\sim a\sqrt{J/D}$, from the injection site, the latter should transform into supercurrent \cite{soninJETP78,*SoninAP10}, as the individual magnons cannot carry spin angular momentum along the $z$ axis. The spin current can suffer relaxation in the ferromagnet, which we account for using Gilbert damping phenomenology. For spin waves, the damping rate at $T\ll T_D$ can be estimated as $\tau^{-1}\sim\al DS/\hbar$, which defines the magnon ballistic decay length $\ell= v\tau$.
We assume $\ell_h\ll\ell$, such that the magnon-mediated current is converted into spin supercurrent without significant decay within the healing length. This imposes a simple condition on Gilbert damping: $\al\ll1$, which is nearly always satisfied in practice.

In order to separate the condensate and magnon contributions to the injected spin current, we parameterize the quantum fields in Eqs.~(\ref{HFI}) and (\ref{exchV}) as a sum of the deterministic classical component (the condensate) and the fluctuating quantum component (magnon cloud):
$\hphi(\bx)=\varphi(x,t)+\de\hphi(\bx)$ and $\hS_z(\bx)=S_{z}(x,t)+\de\hS_z(\bx)$ [with a corresponding decomposition for $\hn_z(\bx)$].
The resultant coupling between the two components affects spin transport both within the ferromagnet bulk and at the interfaces. In the bulk, this coupling manifests only at higher orders in the driving field $\mu_s$~\cite{smf}. At the interfaces, however, the coupling contributes to spin current at linear order in the driving field, as we show below. 

To compute the magnon contribution to spin current through the interface, we insert the above parameterization for $\hphi(\bx)$ and $\hS_z(\bx)$, along with the steady-state ansatz $\varphi(x,t)=\varphi(x)+\W t$ and $S_{z}(x,t)=\hbar\W/Da^3$, into Eq.~(\ref{exchV}).
The precession frequency $\W$ now needs to be self-consistently determined in the presence of the thermal corrections. Since the condensate-magnon coupling only leads to nonlinear effects in the bulk (as argued above), linear-response spin transport should be well-characterized by the condensate described by the above steady-state solution for $\varphi$ and $S_{z}$ together with the decoupled thermal magnon cloud governed by the Hamiltonian $\de\hH_F=\int d^3\bx[A(\nabla\de\hphi)^2+K\de\hn_z^2]/2$.


We first evaluate the condensate contribution to the spin current at each interface. In the absence of the fluctuations, the relevant interaction Hamiltonian is $\hV_0=\sum_r\hV_{r0}$, with
\beq
\label{V0}
\hV_{r0}=\int\frac{d^2\bk}{(2\p)^2}\frac{dk_x}{2\p}\frac{dk_x'}{2\p}\eta_{r0k_xk_x'}e^{i\W t}
\hpsi^\dag_{\bk k_x\up r}\hpsi_{\bk k_x'\down r}+{\rm H.c.}\ ,
\eeq
where $\eta_{r0k_xk_x'}=\eta_r \tS e^{-if(x_r)}\phi^{r*}_{k_x}(x_r)\phi^r_{k_x'}(x_r)/2$.
The reservoir electron operators were expanded as
$\hpsi_{\s r}(\bx)=\int \frac{d^2\bk dk_x}{(2\p)^3}e^{i\bk\cdot\br}\phi^r_{k_x}(x)\hpsi_{\bk k_x\s r}$,
where $\phi^r_{k_x}(x)$ are orthonormal eigenfunctions in the transport direction for the semi-infinite reservoir $r$.
Here, we consider the weak-coupling regime and compute the spin current 
to lowest nontrivial order in $\eta_r$~\cite{weakcoupling}.
In Eq.~(\ref{V0}), we dropped the $z$-component exchange, since it does not contribute to the spin current within the weak-coupling 
treatment. The operator for the spin-current density flowing \textit{into} each reservoir is
\beq
\label{Js0op}
\hJ^s_{r0}=
\frac{i}{2\mathcal{A}}\int\frac{d^2\bk}{(2\p)^2}\frac{dk_x}{2\pi}\sum_\s\square{\hV_{r0},\s\hpsi^\dag_{\bk k_x\s r}\hpsi_{\bk k_x\s r}},
\eeq
where $\mathcal{A}$ is the interface cross-sectional area.
From the Kubo formula, we obtain $J^s_{r0}=-(i/\hbar)\int dt'\thi(-t')\langle[\hJ^s_{r0}(0),\hV_{r0}(t')]\rangle$, where $\theta(t)$ is the Heaviside step function.
To linear order in $\mu_s$ and $\W\propto\mu_s$, we obtain $J^s_{r0}=G^s_{r}\hbar\W_r$, where
$\W_L=\W-\mu_s/\hbar$, and $\W_R=\W$. (In order to consider spin current \textit{into} the ferromagnet \textit{from} the left reservoir, as in Fig.~\ref{setup}, we must flip the sign of $\Omega_L$.)
The conductances read $G^s_{r}=2\p\int_{-\infty}^\infty d\ve\nu_r(\ve)[-n'_{\rm FD}(\ve)]$, where $n'_{\rm FD}(\ve)\equiv\pd n_{\rm FD}(\ve)/\pd\ve$ and
\begin{equation}
\label{nur0}
\nu_r(\ve)=\int\frac{d^2\bk}{(2\p)^2}\frac{dk_x}{2\p}\frac{dk_x'}{2\p}\abs{\eta_{r0k_xk_x'}}^2\de(\ve-\e_{\bk k_x r})\de(\ve-\e_{\bk k_x' r})\,.
\end{equation}

For the magnon contribution, we expand the interaction to linear order in the fluctuations: $\de\hV=\sum_r\de\hV_r$ with
\beq
\label{dV}
\de\hV_{r}=\int_{\{\bk\}}\sum_n^{\hspace{6mm}\prime}\eta_{rn k_xk_x'}e^{-i\W t}\de\hphi_{\bk-\bk'n}\hpsi^\dag_{\bk k_x\up r}\hpsi_{\bk' k_x'\down r}+{\rm H.c.},
\eeq
where $\eta_{rn k_xk_x'}=-i\phi^F_n(x_r)\eta_{r0k_xk_x'}$ and $\int_{\{\bk\}}$ denotes integral over momenta, $\bk$, $\bk'$, $k_x$, and $k'_x$, with the appropriate $(2\p)^{-1}$ factors. Here, we have introduced orthonormal eigenfunctions $\f^F_n(x)=\sqrt{2/L}\cos(q_nx)$
(with non-negative integers $n\ge 0$ and $q_n=n\p/L$), 
which correspond to eigenstates of a free particle in the domain $0\le x\le L$ obeying Neumann boundary conditions, 
$\pd_x\f^F_n(0)=0=\pd_x\f^F_n(L)$. The spin-wave operators are expanded as
$\de\hphi(\bx)=\int \frac{d^2\bq}{(2\p)^2}\sum'_{n}e^{i\bq\cdot\br}\phi^F_n(x)\de\hphi_{\bq n}$, with an analogous expansion for
$\de\hS_z(\bx)$.
We use prime on the summation sign to indicate that it excludes uniform (i.e., condensate) mode
with $\bq=0$ and $n=0$. The magnon current-density operator, $\de\hJ^s_r$, is then given by the
right-hand side of Eq.~(\ref{Js0op}) but with $\hV_{r0}$ replaced by $\de\hV_{r}$. The steady-state magnon 
spin current across the interface is then similarly obtained through 
$\de J^s_{r}=-(i/\hbar)\int dt'\thi(-t')\langle[\de\hJ^s_{r}(0),\de\hV_{r}(t')]\rangle$.

The Hamiltonian for the fluctuations, $\de\hH_F$, can be diagonalized using the ladder operators, $\ha_{\bq n}$
and $\ha^\dag_{\bq n}$, obeying $[\ha_{\bq n},\ha^\dag_{\bq' n'}]=(2\p)^2\de(\bq-\bq')\de_{nn'}$. We 
obtain $\de\hH_F=\int \frac{d^2\bq}{(2\p)^2}\sum^{'}_{n}
E_{\bq n}(\ha^\dag_{\bq n}\ha_{\bq n}+1/2)$, where the magnon spectrum is
$E_{\bq n}=\hbar v\sqrt{|\bq|^2+q_n^2}$. In this basis, the phase field reads
$\de\hphi_{\bq n}=\sqrt{Da^3/2E_{\bq n}}(\ha^\dag_{-\bq n}+\ha_{\bq n})$. 
The linearized thermal contribution to the injected spin-current density reads
$\de J^s_r=\de G^s_r\hbar\W_r$, where the magnon conductances are given by 
$\de G^s_r=4\p\int_0^\infty d\ve\ve\de\nu_{r}(\ve)[-n'_{\rm BE}(\ve)]$, in terms of
\beq
\label{dnumag}
\de\nu_r(\ve)=\hspace{-1.5mm}\int_{\{\bk\}}\hspace{-1.5mm}\sum_n^{\hspace{6mm}\prime}\hspace{-0.5mm}\abs{\eta_{rnk_xk_x'}}^2\hspace{-0.5mm}B_{\bk-\bk' n}(\ve)\de(\mu-\e_{\bk k_xr})\de(\mu-\e_{\bk'k_x'r}),
\eeq
and the magnon spectral function 
$B_{\bq n}(\ve)=(D a^3/2E_{\bq n})\de(\ve-E_{\bq n})$.
Here, $n'_{\rm BE}(\ve)\equiv\pd n_{\rm BE}(\ve)/\pd\ve$, $n_{\rm BE}(\ve)=(e^{\be\ve}-1)^{-1}$, and we assumed that $\mu\gg k_BT$.
The total injected spin current density is then given by $J^s_r= \sG^s_r\hbar\W_r$, where $\sG^s_r=G^s_r+\de G^s_r$. This is a main result of our microscopic calculation.
Due to Gilbert damping in the ferromagnetic bulk, the injected spin current, $J^s_L$, and the collected
spin current, $J^s_R$, are related via $J^s_L-J^s_R=\Delta J^s$, where $\Delta J^s=\al s\W L$ is the condensate spin relaxation in the bulk. Since the total injected spin current is fully transformed into 
supercurrent in the bulk, we thus reproduce Eqs.~\eqref{mainresult}, with the substitution $g_r^{\uparrow\downarrow}\to4\pi \sG^s_r$.

In order to quantify the magnon contribution to the spin transfer with respect to the coherent contribution, we evaluate the ratio $\mathcal{R}\equiv\de G^s_r/G^s_{r}$. In the low-temperature regime considered thus far (i.e., $T\ll T_D,T_c$), we obtain $\mathcal{R}=(\sqrt{S}/6)\sqrt{T/T_D}(T/T_c)^{3/2}$, where $T_c=JS^2/k_B$ and $T_D=DS/k_B$ (see Supplementary Material for details). Here, we see that the magnon contribution to the spin transfer is very small.
In the high-temperature regime, $T_D\ll T\ll T_c$, thermal magnons are no longer strongly affected by the
planar anisotropy and thus acquire a circular character. In this case, the ratio of the magnon
to condensate spin currents becomes $\mathcal{R}^c=(\sqrt{S}/2\pi^2)\G(5/2)\z(3/2)(T/T_c)^{3/2}$, where
$\G$ is the gamma function and $\zeta$ is the Riemann zeta function (see Supplementary Material for details). We see that the magnon contribution remains small as long as $T\ll T_c$.

\begin{figure}[t]
\centering
\includegraphics*[width=0.9\linewidth]{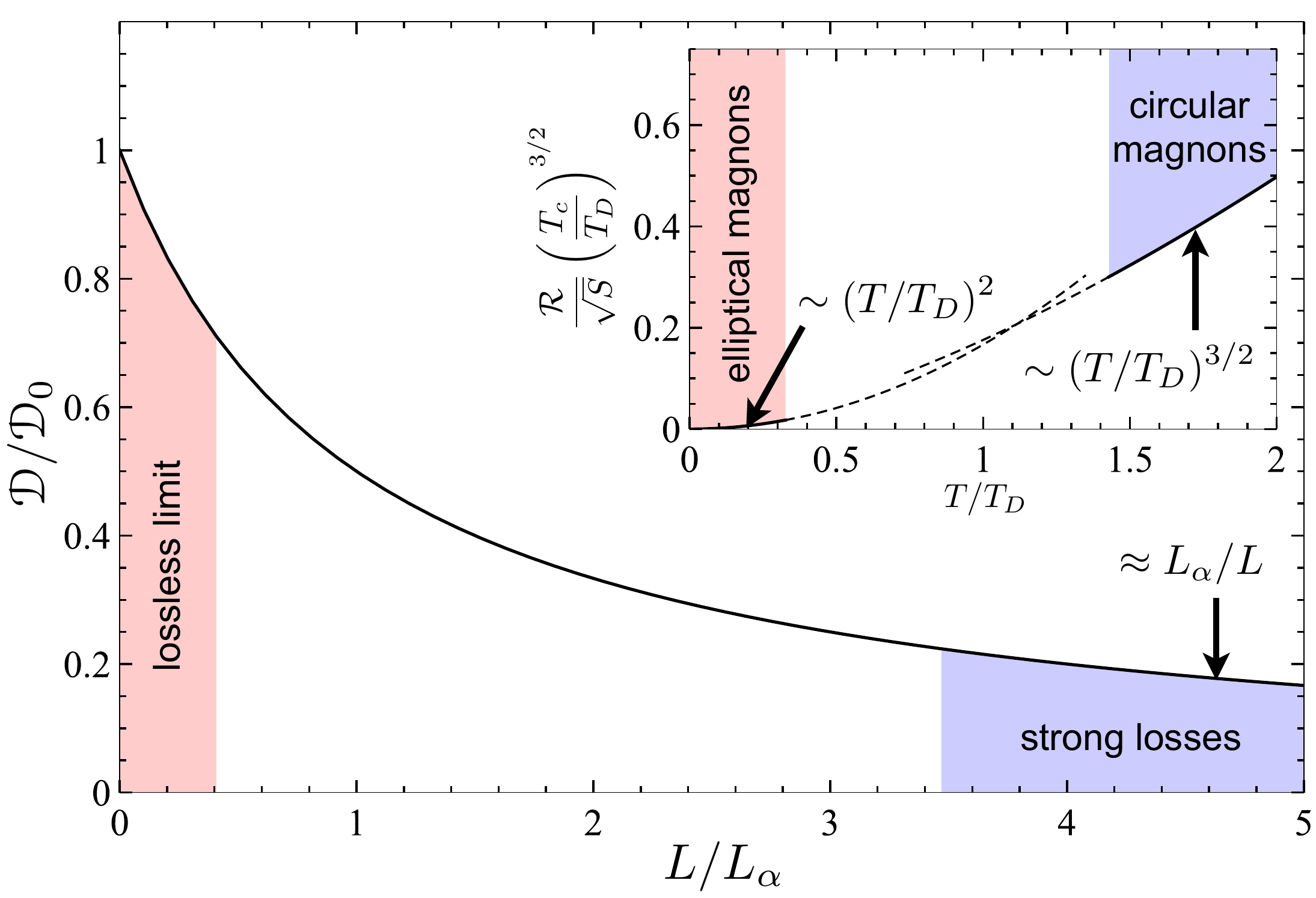}
\caption{Negative drag coefficient normalized by its lossless value $\mathcal{D}_0=\theta_{\rm SH}^2\sigma/2g_Qg^{\uparrow\downarrow}d$ as a function of the length of the ferromagnet $L$. (Inset) Normalized magnon correction $\mathcal{R}$ to the spin-mixing conductance \eqref{R} as a function of the ambient temperature $T$.}
\label{plot}
\end{figure}

{\em Discussion}.|The superfluid spin transport can be detected using the setup shown in Fig.~\ref{setup}. Here, the ferromagnet is sandwiched by identical metals with strong spin-orbit coupling, characterized by an effective spin Hall angle $\theta_{\rm SH}$ at the ferromagnet$|$metal interfaces. Let $J^c_R$ denote the charge-current density produced via the inverse spin Hall effect in the right metal, given the applied charge-current density $J^c_L$ in the left metal, which define the (negative) drag coefficient $\mathcal{D}\equiv -J^c_R/J^c_L$. The spin current impinging on the static ferromagnet at the left interface is given by $J^s_{\rm SH}=(\hbar/2e)\theta_{\rm SH}J^c_L$. Using Onsager reciprocity, the induced charge-current density reads $J_R^c=-(\theta_{\rm SH}\sigma/d)\hbar\Omega/2e$, where $\s$ ($d$) are the conductivity (thickness) of the right metal; here, we assume $d\gtrsim\la_{\rm sf}$, where $\la_{\rm sf}$ is the spin-flip length of the metal. In the absence of magnetic losses and assuming $\theta_{\rm SH}\ll1$, $J^s_R=J^s_L=J^s_{\rm SH}/2$, which gives $\hbar\Omega=(2\pi/g^{\uparrow\downarrow})J^s_{\rm SH}$, resulting in $\mathcal{D}_0=\theta_{\rm SH}^2\sigma/2g_Qg^{\uparrow\downarrow}d$, where $g_Q\equiv2e^2/h$. In the presence of losses, the precession frequency is suppressed according to Eq.~\eqref{mainresult} as $\hbar\W=[4\p/(2g^{\up\down}+g_\al)]J^s_{\rm SH}$, which results in $\mathcal{D}=\mathcal{D}_0/(1+L/L_\al)$, where $L_\al\equiv\hbar g^{\up\down}/2\pi\alpha s$. At finite temperatures, the effective mixing conductance acquires thermal corrections and can be written in the form,
\begin{equation}
g^{\up\down}_{\rm eff}(T)=g^{\up\down}_0(T)[1+\mathcal{R}(T)]\,.
\label{R}
\end{equation}
These results are summarized in Fig.~\ref{plot}. Here, $\mathcal{R}(T)$ denotes a correction to the effective mixing conductance arising from the additional magnon channel for spin transport and $g^{\up\down}_0(T)$ denotes finite temperature interfacial mixing conductance for spin transfer into the condensate. We note that while thermal magnons provide a new channel for transport that gives a positive correction to the effective mixing conductance [as computed above and encoded in $\mathcal{R}(T)$], finite temperature also suppresses the ferromagnetic order and leads to a reduction of the superfluid component of the mixing conductance [as encoded in $g^{\up\down}_0(T)$]. This competition between the two corrections can lead to an {\it overall} suppression in the effective mixing conductance (see Supplementary Material).


For a  quantitative estimate, we consider a Pt$|$YIG$|$Pt hybrid structure (which appears to be a promising combination because of strong spin-orbit coupling in Pt and low Gilbert damping and weak magnetic anisotropy in YIG). Using $\theta_{\rm SH}\sim0.1$ (measured for platinum$|$permalloy interface~\cite{liuPRL11}), $\s\sim0.1~(\mu\W\cdot\mbox{cm})^{-1}$ for Pt, $d\approx\la_{\rm sf}\sim 1$~nm (spin-flip length in Pt~\cite{liuPRL11}), and $g^{\up\down}\sim5\times10^{18}$~m$^{-2}$ for the YIG$|$Pt interfaces~\cite{burrowesAPL12}, we get $\mathcal{D}_0\sim0.1$~\cite{drag}. Taking $\al\sim10^{-4}$ and using YIG spin density $s/\hbar\sim10^{22}~{\rm cm}^{-3}$~\cite{bhagatPSS73}, we get for the crossover length $L_{\alpha}\sim 1~\mu$m. The large and long-ranged negative drag constitute our key predictions.

Finally, we remark that breaking of the U(1) symmetry within the easy-plane of the ferromagnet is detrimental to the DC spin-carrying superfluid state studied here. Relevant macroscopic manifestations of this symmetry breaking are Gilbert damping, which has already been accounted for, and magnetic anisotropy. In the presence of the latter, the applied current must overcome a threshold in order to establish the spin superfluid-carrying state over the length of the ferromagnet. However, the applied current cannot exceed an upper critical current, beyond which the induced planar magnetic spiral state becomes unstable \cite{konigPRL01,soninJETP78,*SoninAP10}.

\acknowledgments

The authors would like to thank Bertrand I. Halperin and Mircea Trif for illuminating discussions. This work was supported in part by FAME (an SRC STARnet center sponsored by MARCO and DARPA), the NSF under Grant No. DMR-0840965, and Grant No. 228481 from the Simons Foundation. This research was supported in part by the Kavli Institute for Theoretical Physics through Grant No. NSF PHY11-25915.


\begin{thebibliography}{36}%
\makeatletter
\providecommand \@ifxundefined [1]{%
 \@ifx{#1\undefined}
}%
\providecommand \@ifnum [1]{%
 \ifnum #1\expandafter \@firstoftwo
 \else \expandafter \@secondoftwo
 \fi
}%
\providecommand \@ifx [1]{%
 \ifx #1\expandafter \@firstoftwo
 \else \expandafter \@secondoftwo
 \fi
}%
\providecommand \natexlab [1]{#1}%
\providecommand \enquote  [1]{``#1''}%
\providecommand \bibnamefont  [1]{#1}%
\providecommand \bibfnamefont [1]{#1}%
\providecommand \citenamefont [1]{#1}%
\providecommand \href@noop [0]{\@secondoftwo}%
\providecommand \href [0]{\begingroup \@sanitize@url \@href}%
\providecommand \@href[1]{\@@startlink{#1}\@@href}%
\providecommand \@@href[1]{\endgroup#1\@@endlink}%
\providecommand \@sanitize@url [0]{\catcode `\\12\catcode `\$12\catcode
  `\&12\catcode `\#12\catcode `\^12\catcode `\_12\catcode `\%12\relax}%
\providecommand \@@startlink[1]{}%
\providecommand \@@endlink[0]{}%
\providecommand \url  [0]{\begingroup\@sanitize@url \@url }%
\providecommand \@url [1]{\endgroup\@href {#1}{\urlprefix }}%
\providecommand \urlprefix  [0]{URL }%
\providecommand \Eprint [0]{\href }%
\providecommand \doibase [0]{http://dx.doi.org/}%
\providecommand \selectlanguage [0]{\@gobble}%
\providecommand \bibinfo  [0]{\@secondoftwo}%
\providecommand \bibfield  [0]{\@secondoftwo}%
\providecommand \translation [1]{[#1]}%
\providecommand \BibitemOpen [0]{}%
\providecommand \bibitemStop [0]{}%
\providecommand \bibitemNoStop [0]{.\EOS\space}%
\providecommand \EOS [0]{\spacefactor3000\relax}%
\providecommand \BibitemShut  [1]{\csname bibitem#1\endcsname}%
\let\auto@bib@innerbib\@empty
\bibitem [{\citenamefont {\v{Z}uti{\'c}}\ \emph {et~al.}(2004)\citenamefont
  {\v{Z}uti{\'c}}, \citenamefont {Fabian},\ and\ \citenamefont {{Das
  Sarma}}}]{zuticRMP04}%
  \BibitemOpen
  \bibfield  {author} {\bibinfo {author} {\bibfnamefont {I.}~\bibnamefont
  {\v{Z}uti{\'c}}}, \bibinfo {author} {\bibfnamefont {J.}~\bibnamefont
  {Fabian}}, \ and\ \bibinfo {author} {\bibfnamefont {S.}~\bibnamefont {{Das
  Sarma}}},\ }\href@noop {} {\bibfield  {journal} {\bibinfo  {journal} {Rev.
  Mod. Phys.}\ }\textbf {\bibinfo {volume} {76}},\ \bibinfo {eid} {323}
  (\bibinfo {year} {2004})}\BibitemShut {NoStop}%
\bibitem [{\citenamefont {Khitun}\ \emph {et~al.}(2008)\citenamefont {Khitun},
  \citenamefont {Bao},\ and\ \citenamefont {Wang}}]{khitunIEEEM08}%
  \BibitemOpen
  \bibfield  {author} {\bibinfo {author} {\bibfnamefont {A.}~\bibnamefont
  {Khitun}}, \bibinfo {author} {\bibfnamefont {M.}~\bibnamefont {Bao}}, \ and\
  \bibinfo {author} {\bibfnamefont {K.~L.}\ \bibnamefont {Wang}},\ }\href@noop
  {} {\bibfield  {journal} {\bibinfo  {journal} {IEEE Trans. Magn.}\ }\textbf
  {\bibinfo {volume} {44}},\ \bibinfo {pages} {2141} (\bibinfo {year}
  {2008})}\BibitemShut {NoStop}%
\bibitem [{\citenamefont {Kajiwara}\ \emph {et~al.}(2010)\citenamefont
  {Kajiwara}, \citenamefont {Harii}, \citenamefont {Takahashi}, \citenamefont
  {Ohe}, \citenamefont {Uchida}, \citenamefont {Mizuguchi}, \citenamefont
  {Umezawa}, \citenamefont {Kawai}, \citenamefont {Ando}, \citenamefont
  {Takanashi}, \citenamefont {Maekawa},\ and\ \citenamefont
  {Saitoh}}]{kajiwaraNAT10}%
  \BibitemOpen
  \bibfield  {author} {\bibinfo {author} {\bibfnamefont {Y.}~\bibnamefont
  {Kajiwara}}, \bibinfo {author} {\bibfnamefont {K.}~\bibnamefont {Harii}},
  \bibinfo {author} {\bibfnamefont {S.}~\bibnamefont {Takahashi}}, \bibinfo
  {author} {\bibfnamefont {J.}~\bibnamefont {Ohe}}, \bibinfo {author}
  {\bibfnamefont {K.}~\bibnamefont {Uchida}}, \bibinfo {author} {\bibfnamefont
  {M.}~\bibnamefont {Mizuguchi}}, \bibinfo {author} {\bibfnamefont
  {H.}~\bibnamefont {Umezawa}}, \bibinfo {author} {\bibfnamefont
  {H.}~\bibnamefont {Kawai}}, \bibinfo {author} {\bibfnamefont
  {K.}~\bibnamefont {Ando}}, \bibinfo {author} {\bibfnamefont {K.}~\bibnamefont
  {Takanashi}}, \bibinfo {author} {\bibfnamefont {S.}~\bibnamefont {Maekawa}},
  \ and\ \bibinfo {author} {\bibfnamefont {E.}~\bibnamefont {Saitoh}},\
  }\href@noop {} {\bibfield  {journal} {\bibinfo  {journal} {Nature}\ }\textbf
  {\bibinfo {volume} {464}},\ \bibinfo {pages} {262} (\bibinfo {year}
  {2010})}\BibitemShut {NoStop}%
\bibitem [{\citenamefont {Uchida}\ \emph {et~al.}(2010)\citenamefont {Uchida},
  \citenamefont {Xiao}, \citenamefont {Adachi}, \citenamefont {Ohe},
  \citenamefont {Takahashi}, \citenamefont {Ieda}, \citenamefont {Ota},
  \citenamefont {Kajiwara}, \citenamefont {Umezawa}, \citenamefont {Kawai},
  \citenamefont {Bauer}, \citenamefont {Maekawa},\ and\ \citenamefont
  {Saitoh}}]{uchidaNATM10}%
  \BibitemOpen
  \bibfield  {author} {\bibinfo {author} {\bibfnamefont {K.}~\bibnamefont
  {Uchida}}, \bibinfo {author} {\bibfnamefont {J.}~\bibnamefont {Xiao}},
  \bibinfo {author} {\bibfnamefont {H.}~\bibnamefont {Adachi}}, \bibinfo
  {author} {\bibfnamefont {J.}~\bibnamefont {Ohe}}, \bibinfo {author}
  {\bibfnamefont {S.}~\bibnamefont {Takahashi}}, \bibinfo {author}
  {\bibfnamefont {J.}~\bibnamefont {Ieda}}, \bibinfo {author} {\bibfnamefont
  {T.}~\bibnamefont {Ota}}, \bibinfo {author} {\bibfnamefont {Y.}~\bibnamefont
  {Kajiwara}}, \bibinfo {author} {\bibfnamefont {H.}~\bibnamefont {Umezawa}},
  \bibinfo {author} {\bibfnamefont {H.}~\bibnamefont {Kawai}}, \bibinfo
  {author} {\bibfnamefont {G.~E.~W.}\ \bibnamefont {Bauer}}, \bibinfo {author}
  {\bibfnamefont {S.}~\bibnamefont {Maekawa}}, \ and\ \bibinfo {author}
  {\bibfnamefont {E.}~\bibnamefont {Saitoh}},\ }\href@noop {} {\bibfield
  {journal} {\bibinfo  {journal} {Nature Mater.}\ }\textbf {\bibinfo {volume}
  {9}},\ \bibinfo {pages} {894} (\bibinfo {year} {2010})}\BibitemShut {NoStop}%
\bibitem [{\citenamefont {Nikuni}\ \emph {et~al.}(2000)\citenamefont {Nikuni},
  \citenamefont {Oshikawa}, \citenamefont {Oosawa},\ and\ \citenamefont
  {Tanaka}}]{nikuniPRL00}%
  \BibitemOpen
  \bibfield  {author} {\bibinfo {author} {\bibfnamefont {T.}~\bibnamefont
  {Nikuni}}, \bibinfo {author} {\bibfnamefont {M.}~\bibnamefont {Oshikawa}},
  \bibinfo {author} {\bibfnamefont {A.}~\bibnamefont {Oosawa}}, \ and\ \bibinfo
  {author} {\bibfnamefont {H.}~\bibnamefont {Tanaka}},\ }\href {\doibase
  10.1103/PhysRevLett.84.5868} {\bibfield  {journal} {\bibinfo  {journal}
  {Phys. Rev. Lett.}\ }\textbf {\bibinfo {volume} {84}},\ \bibinfo {pages}
  {5868} (\bibinfo {year} {2000})}\BibitemShut {NoStop}%
\bibitem [{\citenamefont {Oosawa}\ \emph {et~al.}(1999)\citenamefont {Oosawa},
  \citenamefont {Ishii},\ and\ \citenamefont {Tanaka}}]{oosawaJPCM99}%
  \BibitemOpen
  \bibfield  {author} {\bibinfo {author} {\bibfnamefont {A.}~\bibnamefont
  {Oosawa}}, \bibinfo {author} {\bibfnamefont {M.}~\bibnamefont {Ishii}}, \
  and\ \bibinfo {author} {\bibfnamefont {H.}~\bibnamefont {Tanaka}},\
  }\href@noop {} {\bibfield  {journal} {\bibinfo  {journal} {J. Phys.: Condens.
  Matter}\ }\textbf {\bibinfo {volume} {11}},\ \bibinfo {pages} {265} (\bibinfo
  {year} {1999})}\BibitemShut {NoStop}%
\bibitem [{\citenamefont {Radu}\ \emph {et~al.}(2005)\citenamefont {Radu},
  \citenamefont {Wilhelm}, \citenamefont {Yushankhai}, \citenamefont
  {Kovrizhin}, \citenamefont {Coldea}, \citenamefont {Tylczynski},
  \citenamefont {L{\"u}hmann},\ and\ \citenamefont {Steglich}}]{raduPRL05}%
  \BibitemOpen
  \bibfield  {author} {\bibinfo {author} {\bibfnamefont {T.}~\bibnamefont
  {Radu}}, \bibinfo {author} {\bibfnamefont {H.}~\bibnamefont {Wilhelm}},
  \bibinfo {author} {\bibfnamefont {V.}~\bibnamefont {Yushankhai}}, \bibinfo
  {author} {\bibfnamefont {D.}~\bibnamefont {Kovrizhin}}, \bibinfo {author}
  {\bibfnamefont {R.}~\bibnamefont {Coldea}}, \bibinfo {author} {\bibfnamefont
  {Z.}~\bibnamefont {Tylczynski}}, \bibinfo {author} {\bibfnamefont
  {T.}~\bibnamefont {L{\"u}hmann}}, \ and\ \bibinfo {author} {\bibfnamefont
  {F.}~\bibnamefont {Steglich}},\ }\href {\doibase
  10.1103/PhysRevLett.95.127202} {\bibfield  {journal} {\bibinfo  {journal}
  {Phys. Rev. Lett.}\ }\textbf {\bibinfo {volume} {95}},\ \bibinfo {pages}
  {127202} (\bibinfo {year} {2005})}\BibitemShut {NoStop}%
\bibitem [{\citenamefont {Demokritov}\ \emph {et~al.}(2006)\citenamefont
  {Demokritov}, \citenamefont {Demidov}, \citenamefont {Dzyapko}, \citenamefont
  {Melkov}, \citenamefont {Serga}, \citenamefont {Hillebrands},\ and\
  \citenamefont {Slavin}}]{demokritovNAT06}%
  \BibitemOpen
  \bibfield  {author} {\bibinfo {author} {\bibfnamefont {S.~O.}\ \bibnamefont
  {Demokritov}}, \bibinfo {author} {\bibfnamefont {V.~E.}\ \bibnamefont
  {Demidov}}, \bibinfo {author} {\bibfnamefont {O.}~\bibnamefont {Dzyapko}},
  \bibinfo {author} {\bibfnamefont {G.~A.}\ \bibnamefont {Melkov}}, \bibinfo
  {author} {\bibfnamefont {A.~A.}\ \bibnamefont {Serga}}, \bibinfo {author}
  {\bibfnamefont {B.}~\bibnamefont {Hillebrands}}, \ and\ \bibinfo {author}
  {\bibfnamefont {A.~N.}\ \bibnamefont {Slavin}},\ }\href@noop {} {\bibfield
  {journal} {\bibinfo  {journal} {Nature}\ }\textbf {\bibinfo {volume} {443}},\
  \bibinfo {pages} {430} (\bibinfo {year} {2006})}\BibitemShut {NoStop}%
\bibitem [{\citenamefont {Halperin}\ and\ \citenamefont
  {Hohenberg}(1969)}]{halperinPR69}%
  \BibitemOpen
  \bibfield  {author} {\bibinfo {author} {\bibfnamefont {B.~I.}\ \bibnamefont
  {Halperin}}\ and\ \bibinfo {author} {\bibfnamefont {P.~C.}\ \bibnamefont
  {Hohenberg}},\ }\href@noop {} {\bibfield  {journal} {\bibinfo  {journal}
  {Phys. Rev.}\ }\textbf {\bibinfo {volume} {188}},\ \bibinfo {pages} {898}
  (\bibinfo {year} {1969})}\BibitemShut {NoStop}%
\bibitem [{\citenamefont {Corruccini}\ and\ \citenamefont
  {Osheroff}(1975)}]{corrucciniPRL75}%
  \BibitemOpen
  \bibfield  {author} {\bibinfo {author} {\bibfnamefont {L.~R.}\ \bibnamefont
  {Corruccini}}\ and\ \bibinfo {author} {\bibfnamefont {D.~D.}\ \bibnamefont
  {Osheroff}},\ }\href {\doibase 10.1103/PhysRevLett.34.564} {\bibfield
  {journal} {\bibinfo  {journal} {Phys. Rev. Lett.}\ }\textbf {\bibinfo
  {volume} {34}},\ \bibinfo {pages} {564} (\bibinfo {year} {1975})}\BibitemShut
  {NoStop}%
\bibitem [{\citenamefont {Vuorio}(1974)}]{vuorioJPC74}%
  \BibitemOpen
  \bibfield  {author} {\bibinfo {author} {\bibfnamefont {M.}~\bibnamefont
  {Vuorio}},\ }\href@noop {} {\bibfield  {journal} {\bibinfo  {journal} {J.
  Phys. C: Sol. State Phys.}\ }\textbf {\bibinfo {volume} {7}},\ \bibinfo
  {pages} {L5} (\bibinfo {year} {1974})}\BibitemShut {NoStop}%
\bibitem [{\citenamefont {Vuorio}(1976)}]{vuorioJPC76}%
  \BibitemOpen
  \bibfield  {author} {\bibinfo {author} {\bibfnamefont {M.}~\bibnamefont
  {Vuorio}},\ }\href@noop {} {\bibfield  {journal} {\bibinfo  {journal} {J.
  Phys. C: Sol. State Phys.}\ }\textbf {\bibinfo {volume} {9}},\ \bibinfo
  {pages} {L267} (\bibinfo {year} {1976})}\BibitemShut {NoStop}%
\bibitem [{\citenamefont {Sonin}(1979)}]{soninJETPL79}%
  \BibitemOpen
  \bibfield  {author} {\bibinfo {author} {\bibfnamefont {E.~B.}\ \bibnamefont
  {Sonin}},\ }\href@noop {} {\bibfield  {journal} {\bibinfo  {journal} {JETP
  Lett.}\ }\textbf {\bibinfo {volume} {30}},\ \bibinfo {pages} {662} (\bibinfo
  {year} {1979})}\BibitemShut {NoStop}%
\bibitem [{\citenamefont {Borovik-Romanov}\ \emph {et~al.}(1984)\citenamefont
  {Borovik-Romanov}, \citenamefont {Bunkov}, \citenamefont {Dmitriev},\ and\
  \citenamefont {Mukharskii}}]{borovikJETPL84}%
  \BibitemOpen
  \bibfield  {author} {\bibinfo {author} {\bibfnamefont {A.~S.}\ \bibnamefont
  {Borovik-Romanov}}, \bibinfo {author} {\bibfnamefont {Y.~M.}\ \bibnamefont
  {Bunkov}}, \bibinfo {author} {\bibfnamefont {V.~V.}\ \bibnamefont
  {Dmitriev}}, \ and\ \bibinfo {author} {\bibfnamefont {Y.~M.}\ \bibnamefont
  {Mukharskii}},\ }\href@noop {} {\bibfield  {journal} {\bibinfo  {journal}
  {JETP Lett.}\ }\textbf {\bibinfo {volume} {40}},\ \bibinfo {pages} {1033}
  (\bibinfo {year} {1984})}\BibitemShut {NoStop}%
\bibitem [{\citenamefont {Fomin}(1984)}]{fominJETPL84}%
  \BibitemOpen
  \bibfield  {author} {\bibinfo {author} {\bibfnamefont {I.~A.}\ \bibnamefont
  {Fomin}},\ }\href@noop {} {\bibfield  {journal} {\bibinfo  {journal} {JETP
  Lett.}\ }\textbf {\bibinfo {volume} {40}},\ \bibinfo {pages} {1037} (\bibinfo
  {year} {1984})}\BibitemShut {NoStop}%
\bibitem [{\citenamefont {Sonin}(1978{\natexlab{a}})}]{soninSSC78}%
  \BibitemOpen
  \bibfield  {author} {\bibinfo {author} {\bibfnamefont {E.~B.}\ \bibnamefont
  {Sonin}},\ }\href@noop {} {\bibfield  {journal} {\bibinfo  {journal} {Solid
  State Commun.}\ }\textbf {\bibinfo {volume} {25}},\ \bibinfo {pages} {253}
  (\bibinfo {year} {1978}{\natexlab{a}})}\BibitemShut {NoStop}%
\bibitem [{\citenamefont {Guseinov}\ and\ \citenamefont
  {Keldysh}(1973)}]{guseinovJETPL73}%
  \BibitemOpen
  \bibfield  {author} {\bibinfo {author} {\bibfnamefont {R.~R.}\ \bibnamefont
  {Guseinov}}\ and\ \bibinfo {author} {\bibfnamefont {L.~V.}\ \bibnamefont
  {Keldysh}},\ }\href@noop {} {\bibfield  {journal} {\bibinfo  {journal} {JETP
  Lett.}\ }\textbf {\bibinfo {volume} {36}},\ \bibinfo {pages} {1193} (\bibinfo
  {year} {1973})}\BibitemShut {NoStop}%
\bibitem [{\citenamefont {K{\"o}nig}\ \emph {et~al.}(2001)\citenamefont
  {K{\"o}nig}, \citenamefont {B{\o}nsager},\ and\ \citenamefont
  {MacDonald}}]{konigPRL01}%
  \BibitemOpen
  \bibfield  {author} {\bibinfo {author} {\bibfnamefont {J.}~\bibnamefont
  {K{\"o}nig}}, \bibinfo {author} {\bibfnamefont {M.~C.}\ \bibnamefont
  {B{\o}nsager}}, \ and\ \bibinfo {author} {\bibfnamefont {A.~H.}\ \bibnamefont
  {MacDonald}},\ }\href@noop {} {\bibfield  {journal} {\bibinfo  {journal}
  {Phys. Rev. Lett.}\ }\textbf {\bibinfo {volume} {87}},\ \bibinfo {pages}
  {187202} (\bibinfo {year} {2001})}\BibitemShut {NoStop}%
\bibitem [{\citenamefont {Murakami}\ \emph {et~al.}(2003)\citenamefont
  {Murakami}, \citenamefont {Nagaosa},\ and\ \citenamefont
  {Zhang}}]{murakamiSCI03}%
  \BibitemOpen
  \bibfield  {author} {\bibinfo {author} {\bibfnamefont {S.}~\bibnamefont
  {Murakami}}, \bibinfo {author} {\bibfnamefont {N.}~\bibnamefont {Nagaosa}}, \
  and\ \bibinfo {author} {\bibfnamefont {S.-C.}\ \bibnamefont {Zhang}},\
  }\href@noop {} {\bibfield  {journal} {\bibinfo  {journal} {Science}\ }\textbf
  {\bibinfo {volume} {301}},\ \bibinfo {pages} {1348} (\bibinfo {year}
  {2003})}\BibitemShut {NoStop}%
\bibitem [{\citenamefont {Sinova}\ \emph {et~al.}(2004)\citenamefont {Sinova},
  \citenamefont {Culcer}, \citenamefont {Niu}, \citenamefont {Sinitsyn},
  \citenamefont {Jungwirth},\ and\ \citenamefont {MacDonald}}]{sinovaPRL04}%
  \BibitemOpen
  \bibfield  {author} {\bibinfo {author} {\bibfnamefont {J.}~\bibnamefont
  {Sinova}}, \bibinfo {author} {\bibfnamefont {D.}~\bibnamefont {Culcer}},
  \bibinfo {author} {\bibfnamefont {Q.}~\bibnamefont {Niu}}, \bibinfo {author}
  {\bibfnamefont {N.~A.}\ \bibnamefont {Sinitsyn}}, \bibinfo {author}
  {\bibfnamefont {T.}~\bibnamefont {Jungwirth}}, \ and\ \bibinfo {author}
  {\bibfnamefont {A.~H.}\ \bibnamefont {MacDonald}},\ }\href@noop {} {\bibfield
   {journal} {\bibinfo  {journal} {Phys. Rev. Lett.}\ }\textbf {\bibinfo
  {volume} {92}},\ \bibinfo {eid} {126603} (\bibinfo {year}
  {2004})}\BibitemShut {NoStop}%
\bibitem [{\citenamefont {Tserkovnyak}\ \emph {et~al.}(2002)\citenamefont
  {Tserkovnyak}, \citenamefont {Brataas},\ and\ \citenamefont
  {Bauer}}]{tserkovPRL02sp}%
  \BibitemOpen
  \bibfield  {author} {\bibinfo {author} {\bibfnamefont {Y.}~\bibnamefont
  {Tserkovnyak}}, \bibinfo {author} {\bibfnamefont {A.}~\bibnamefont
  {Brataas}}, \ and\ \bibinfo {author} {\bibfnamefont {G.~E.~W.}\ \bibnamefont
  {Bauer}},\ }\href@noop {} {\bibfield  {journal} {\bibinfo  {journal} {Phys.
  Rev. Lett.}\ }\textbf {\bibinfo {volume} {88}},\ \bibinfo {eid} {117601}
  (\bibinfo {year} {2002})}\BibitemShut {NoStop}%
\bibitem [{\citenamefont {Tserkovnyak}\ \emph {et~al.}(2005)\citenamefont
  {Tserkovnyak}, \citenamefont {Brataas}, \citenamefont {Bauer},\ and\
  \citenamefont {Halperin}}]{tserkovRMP05}%
  \BibitemOpen
  \bibfield  {author} {\bibinfo {author} {\bibfnamefont {Y.}~\bibnamefont
  {Tserkovnyak}}, \bibinfo {author} {\bibfnamefont {A.}~\bibnamefont
  {Brataas}}, \bibinfo {author} {\bibfnamefont {G.~E.~W.}\ \bibnamefont
  {Bauer}}, \ and\ \bibinfo {author} {\bibfnamefont {B.~I.}\ \bibnamefont
  {Halperin}},\ }\href@noop {} {\bibfield  {journal} {\bibinfo  {journal} {Rev.
  Mod. Phys.}\ }\textbf {\bibinfo {volume} {77}},\ \bibinfo {eid} {1375}
  (\bibinfo {year} {2005})}\BibitemShut {NoStop}%
\bibitem [{\citenamefont {Slonczewski}(1996)}]{slonczewskiJMMM96}%
  \BibitemOpen
  \bibfield  {author} {\bibinfo {author} {\bibfnamefont {J.~C.}\ \bibnamefont
  {Slonczewski}},\ }\href@noop {} {\bibfield  {journal} {\bibinfo  {journal}
  {J. Magn. Magn. Mater.}\ }\textbf {\bibinfo {volume} {159}},\ \bibinfo
  {pages} {L1} (\bibinfo {year} {1996})}\BibitemShut {NoStop}%
\bibitem [{\citenamefont {Zhang}\ and\ \citenamefont
  {Zhang}(2012)}]{zhangPRL12}%
  \BibitemOpen
  \bibfield  {author} {\bibinfo {author} {\bibfnamefont {S.~S.-L.}\
  \bibnamefont {Zhang}}\ and\ \bibinfo {author} {\bibfnamefont
  {S.}~\bibnamefont {Zhang}},\ }\href {\doibase 10.1103/PhysRevLett.109.096603}
  {\bibfield  {journal} {\bibinfo  {journal} {Phys. Rev. Lett.}\ }\textbf
  {\bibinfo {volume} {109}},\ \bibinfo {pages} {096603} (\bibinfo {year}
  {2012})}\BibitemShut {NoStop}%
\bibitem [{\citenamefont {Sonin}(1978{\natexlab{b}})}]{soninJETP78}%
  \BibitemOpen
  \bibfield  {author} {\bibinfo {author} {\bibfnamefont {E.~B.}\ \bibnamefont
  {Sonin}},\ }\href@noop {} {\bibfield  {journal} {\bibinfo  {journal} {Sov. Phys. JETP}\ }\textbf {\bibinfo {volume} {47}},\ \bibinfo {pages} {1091} (\bibinfo
  {year} {1978}{\natexlab{b}})}\BibitemShut {NoStop}%
\bibitem [{\citenamefont {Sonin}(2010)}]{SoninAP10}%
  \BibitemOpen
  \bibfield  {author} {\bibinfo {author} {\bibfnamefont {E.~B.}\ \bibnamefont
  {Sonin}},\ }\href@noop {} {\bibfield  {journal} {\bibinfo  {journal} {Adv.
  Phys.}\ }\textbf {\bibinfo {volume} {59}},\ \bibinfo {pages} {181} (\bibinfo
  {year} {2010})}\BibitemShut {NoStop}%
\bibitem [{smf()}]{smf}%
  \BibitemOpen
  \href@noop {} {}\bibinfo {howpublished} {In the high-temperature
  (circular-magnon) regime, $T_D\ll T$, the bulk coupling between a
  long-wavelength condensate dynamics and thermal cloud acquires a particularly
  simple form. In this case, magnons experience a reactive and dissipative
  spin-motive forces, whose Cartesian components are proportional to
  $\bn\cdot(\pd_t\bn\times\nabla_i\bn)$ \cite{volovikJPC87} and
  $\pd_t\bn\cdot\nabla_i\bn$ \cite{DuineetalPRB2008,*TserkovnyaketalPRB2008},
  respectively. Since both $\pd_t\bn$ and $\nabla_i\bn$ are linear in the
  nonequilibrium drive, the effects of these forces would appear only at
  quadratic order.}\BibitemShut {Stop}%
\bibitem [{wea()}]{weakcoupling}%
  \BibitemOpen
  \href@noop {} {}\bibinfo {howpublished} {By weak coupling regime can be
  justified here if the energy scale associated with the interface exchange
  coupling, i.e. $\eta_r/a^4$ is much smaller than the typical magnon energy
  scale, $\ve_m\sim\mbox{max}\{k_BT, DS\}$ or the typical energy scale of the
  conduction electrons in the reservoir, which is the Fermi energy $\ve_F$.
  Since we expect $\ve_m\ll\ve_F$, the weak-coupling treatment should be valid
  for $\eta_r/a^4\ll\ve_m$.}\BibitemShut {Stop}%
\bibitem [{\citenamefont {Liu}\ \emph {et~al.}(2011)\citenamefont {Liu},
  \citenamefont {Bian}, \citenamefont {Miller},\ and\ \citenamefont
  {Chiang}}]{liuPRL11}%
  \BibitemOpen
  \bibfield  {author} {\bibinfo {author} {\bibfnamefont {L.}~\bibnamefont
  {Liu}}, \bibinfo {author} {\bibfnamefont {T.}~\bibnamefont {Moriyama}}, \bibinfo
  {author} {\bibfnamefont {D. C.}~\bibnamefont {Ralph}}, and\ \bibinfo {author}
  {\bibfnamefont {R. A.}\ \bibnamefont {Buhrman}},\ }\href {\doibase
  10.1103/PhysRevLett.106.036601} {\bibfield  {journal} {\bibinfo  {journal}
  {Phys. Rev. Lett.}\ }\textbf {\bibinfo {volume} {106}},\ \bibinfo {pages}
  {036601} (\bibinfo {year} {2011})}\BibitemShut {NoStop}%
\bibitem [{\citenamefont {Burrowes}\ \emph {et~al.}(2012)\citenamefont
  {Burrowes}, \citenamefont {Heinrich}, \citenamefont {Kardasz}, \citenamefont
  {Montoya}, \citenamefont {Girt}, \citenamefont {Sun}, \citenamefont {Song},\
  and\ \citenamefont {Wu}}]{burrowesAPL12}%
  \BibitemOpen
  \bibfield  {author} {\bibinfo {author} {\bibfnamefont {C.}~\bibnamefont
  {Burrowes}}, \bibinfo {author} {\bibfnamefont {B.}~\bibnamefont {Heinrich}},
  \bibinfo {author} {\bibfnamefont {B.}~\bibnamefont {Kardasz}}, \bibinfo
  {author} {\bibfnamefont {E.~A.}\ \bibnamefont {Montoya}}, \bibinfo {author}
  {\bibfnamefont {E.}~\bibnamefont {Girt}}, \bibinfo {author} {\bibfnamefont
  {Y.}~\bibnamefont {Sun}}, \bibinfo {author} {\bibfnamefont {Y.-Y.}\
  \bibnamefont {Song}}, \ and\ \bibinfo {author} {\bibfnamefont
  {M.}~\bibnamefont {Wu}},\ }\href {\doibase 10.1063/1.3690918} {\bibfield
  {journal} {\bibinfo  {journal} {Appl. Phys. Lett.}\ }\textbf {\bibinfo
  {volume} {100}},\ \bibinfo {eid} {092403} (\bibinfo {year}
  {2012})}\BibitemShut {NoStop}%
\bibitem [{dra()}]{drag}%
  \BibitemOpen
  \href@noop {} {}\bibinfo {howpublished} {For purely magnon-mediated drag in a
  Pt$|$YIG$|$Pt hybrid structure, the drag coefficient has been theoretically
  proposed and estimated to be $\sim10^{-4}$ \cite{zhangPRL12}. Ref.~\cite{kajiwaraNAT10} reported
  a drag $\sim10^{-9}$ in a Pt$|$YIG$|$Pt system (which was interpreted to develop as a result of a nonlinear magnon
  generation).}\BibitemShut {Stop}%
\bibitem [{\citenamefont {Bhagat}\ \emph {et~al.}(1973)\citenamefont {Bhagat},
  \citenamefont {Lessoff}, \citenamefont {Vittoria},\ and\ \citenamefont
  {Guenzer}}]{bhagatPSS73}%
  \BibitemOpen
  \bibfield  {author} {\bibinfo {author} {\bibfnamefont {S.}~\bibnamefont
  {Bhagat}}, \bibinfo {author} {\bibfnamefont {H.}~\bibnamefont {Lessoff}},
  \bibinfo {author} {\bibfnamefont {C.}~\bibnamefont {Vittoria}}, \ and\
  \bibinfo {author} {\bibfnamefont {C.}~\bibnamefont {Guenzer}},\ }\href@noop
  {} {\bibfield  {journal} {\bibinfo  {journal} {Phys. Status Solidi}\ }\textbf
  {\bibinfo {volume} {20}},\ \bibinfo {pages} {731} (\bibinfo {year}
  {1973})}\BibitemShut {NoStop}%
\bibitem [{\citenamefont {Volovik}(1987)}]{volovikJPC87}%
  \BibitemOpen
  \bibfield  {author} {\bibinfo {author} {\bibfnamefont {G.~E.}\ \bibnamefont
  {Volovik}},\ }\href@noop {} {\bibfield  {journal} {\bibinfo  {journal} {J.
  Phys. C: Sol. State Phys.}\ }\textbf {\bibinfo {volume} {20}},\ \bibinfo
  {pages} {L83} (\bibinfo {year} {1987})}\BibitemShut {NoStop}%
\bibitem [{\citenamefont {Duine}(2008)}]{DuineetalPRB2008}%
  \BibitemOpen
  \bibfield  {author} {\bibinfo {author} {\bibfnamefont {R.~A.}\ \bibnamefont
  {Duine}},\ }\href {\doibase 10.1103/PhysRevB.77.014409} {\bibfield  {journal}
  {\bibinfo  {journal} {Phys. Rev. B}\ }\textbf {\bibinfo {volume} {77}},\
  \bibinfo {pages} {014409} (\bibinfo {year} {2008})}\BibitemShut {NoStop}%
\bibitem [{\citenamefont {Tserkovnyak}\ and\ \citenamefont
  {Mecklenburg}(2008)}]{TserkovnyaketalPRB2008}%
  \BibitemOpen
  \bibfield  {author} {\bibinfo {author} {\bibfnamefont {Y.}~\bibnamefont
  {Tserkovnyak}}\ and\ \bibinfo {author} {\bibfnamefont {M.}~\bibnamefont
  {Mecklenburg}},\ }\href {\doibase 10.1103/PhysRevB.77.134407} {\bibfield
  {journal} {\bibinfo  {journal} {Phys. Rev. B}\ }\textbf {\bibinfo {volume}
  {77}},\ \bibinfo {pages} {134407} (\bibinfo {year} {2008})}\BibitemShut
  {NoStop}%
\end{thebibliography}

\begin{thebibliography}{35}%
\makeatletter
\providecommand \@ifxundefined [1]{%
 \@ifx{#1\undefined}
}%
\providecommand \@ifnum [1]{%
 \ifnum #1\expandafter \@firstoftwo
 \else \expandafter \@secondoftwo
 \fi
}%
\providecommand \@ifx [1]{%
 \ifx #1\expandafter \@firstoftwo
 \else \expandafter \@secondoftwo
 \fi
}%
\providecommand \natexlab [1]{#1}%
\providecommand \enquote  [1]{``#1''}%
\providecommand \bibnamefont  [1]{#1}%
\providecommand \bibfnamefont [1]{#1}%
\providecommand \citenamefont [1]{#1}%
\providecommand \href@noop [0]{\@secondoftwo}%
\providecommand \href [0]{\begingroup \@sanitize@url \@href}%
\providecommand \@href[1]{\@@startlink{#1}\@@href}%
\providecommand \@@href[1]{\endgroup#1\@@endlink}%
\providecommand \@sanitize@url [0]{\catcode `\\12\catcode `\$12\catcode
  `\&12\catcode `\#12\catcode `\^12\catcode `\_12\catcode `\%12\relax}%
\providecommand \@@startlink[1]{}%
\providecommand \@@endlink[0]{}%
\providecommand \url  [0]{\begingroup\@sanitize@url \@url }%
\providecommand \@url [1]{\endgroup\@href {#1}{\urlprefix }}%
\providecommand \urlprefix  [0]{URL }%
\providecommand \Eprint [0]{\href }%
\providecommand \doibase [0]{http://dx.doi.org/}%
\providecommand \selectlanguage [0]{\@gobble}%
\providecommand \bibinfo  [0]{\@secondoftwo}%
\providecommand \bibfield  [0]{\@secondoftwo}%
\providecommand \translation [1]{[#1]}%
\providecommand \BibitemOpen [0]{}%
\providecommand \bibitemStop [0]{}%
\providecommand \bibitemNoStop [0]{.\EOS\space}%
\providecommand \EOS [0]{\spacefactor3000\relax}%
\providecommand \BibitemShut  [1]{\csname bibitem#1\endcsname}%
\let\auto@bib@innerbib\@empty
\bibitem [{\citenamefont {Datta}(1995)}]{dattaBOOK95}%
  \BibitemOpen
  \bibfield  {author} {\bibinfo {author} {\bibfnamefont {S.}~\bibnamefont
  {Datta}},\ }\href@noop {} {\emph {\bibinfo {title} {Electronic Transport in
  Mesoscopic Systems}}}\ (\bibinfo  {publisher} {Cambridge University Press},\
  \bibinfo {address} {Cambridge},\ \bibinfo {year} {1995})\BibitemShut
  {NoStop}%
\end{thebibliography}
%

\newpage
\appendix
\section{Supplementary Material}

In this supplementary material, we first discuss how the ratios $\mathcal{R}$ and $\mathcal{R}^c$ are obtained. We then briefly study spin transport when the spin accumulation is oriented parallel to the ferromagnet's easy-plane.\\

{\em Elliptical magnons}.|We begin in the low temperature regime, $T\ll T_D,T_c$, where the magnons have an elliptical character due to the easy-plane anisotropy.
Here, we compute $\mathcal{R}$ to lowest order in temperature while evaluating $G^s_{r}$ at zero temperature [where $G^s_r|_{T=0}=2\p\nu_r(\mu)]$.
We assume that the thermal-magnon
wavelength $\lambda_T\gg\lambda_F$, where $\lambda_F$ is the electron Fermi wavelength in the reservoirs. This allows us to factor Eq.~(9) in the main text into an electronic contribution of the form (7) in the main text and a magnonic contribution,
\beq
\label{denu}
\de\nu_r(\ve)\approx\nu_r(\mu)\frac{2}{L}\sum^{\hspace{6mm}\prime}_n\int\frac{d^2\bq}{(2\p)^2}B_{\bq n}(\ve),
\eeq
where we have used $|\f^F_n(x_r)|^2=2/L$. Replacing the above summation by a momentum integral and using the magnon spectral function appropriate for the elliptical regime (as defined in the main text),
we obtain $\mathcal{R}=(\sqrt{S}/6)\sqrt{T/T_D}(T/T_c)^{3/2}$, where $T_c$ and $T_D$ have been defined in the main text.\\


{\em Circular magnons}.|For $T_D\ll T\ll T_c$, thermal magnons are no longer strongly affected by the
planar anisotropy and thus acquire a circular character. The magnon component to the total spin conductance is 
still given by $\de G^s_r$, according to Eq.~(9) in the main text, but the spectral function is now modified to
$B^c_{\bq n}(\ve)=\de(\ve-E^c_{\bq n})/2\tS$, with $E^c_{\bq n}=JS a^2 (|\bq|^2+q_n^2)$.
We remark here that the healing length and the damping rate are given by $\ell_h\sim a\sqrt{JS/k_BT}$
and $\tau^{-1}\sim\al k_BT/\hbar$, respectively, and the corresponding ballistic decay length is $\ell=v\tau$, where the
thermal-magnon velocity now reads $v\sim\sqrt{JSk_BT}a/\hbar$. The assumption that the magnon healing length is much less than the magnon ballistic decay length (efficient magnon-to-condensate conversion near the spin current injection site), i.e. $\ell\gg\ell_h$, translates again to $\al\ll1$. Again assuming that the thermal-magnon wavelength $\lambda_T\gg\lambda_F$, Eq.~(9) in the main text can be factored into the form given in Eq.~\eqref{denu} above as for the elliptical case, but with $B_{\bq n}(\ve)$ replaced by the magnon spectral function appropriate for the circular regime, $B^c_{\bq n}(\ve)$. The ratio of the thermal to condensate spin currents (while again evaluating $G^s_{r}$ at zero temperature) then becomes $\mathcal{R}^c=(2/\tS)\int_0^\infty d\ve\rho^c_m(\ve)\ve[-n'_{\rm BE}(\ve)]=(\sqrt{S}/2\pi^2)\G(5/2)\z(3/2)(T/T_c)^{3/2}$, where $\rho^c_m(\ve)=\sqrt{\ve}/4\p^2(JS)^{3/2}a^3$ is the magnon density of states, $\G$ is the gamma function and $\zeta$ is the Riemann zeta function.\\

{\em Thermal correction to the superfluid conductance}.|We note here that the superfluid conductance $G^s_r$ also has a thermal correction that is of the same order as the correction due to the magnon channel considered above. At finite temperatures, the ferromagnetic order parameter is reduced and, consequently, leads to a suppression of the superfluid component of the interfacial spin transfer. For the circular magnon case, we have verified that this correction leads to an {\rm overall} suppression of the effective interfacial spin-mixing conductance, i.e., $g^{\up\down}_{\rm eff}(T)\approx g^{\up\down}_0(T)(1+\mathcal{R}^c)$, where $\mathcal{R}^c$ is the correction calculated above, $g^{\up\down}_0(T)=g^{\up\down}_{\rm eff}(T=0)(1-2n/\tilde S)$, and $2n/\tilde S=(4/3)\mathcal{R}^c$ ($n$ being the thermal magnon density) is the correction arising due to the reduced order parameter. The appearance of the overall suppression factor $(1-2n/\tilde S)$ is a consequence of the replacement $\tilde S\rightarrow\tilde S-n$ in the matrix element in Eq.~(7) of the main text and evaluating the (square of the) matrix element to lowest order in $n/\tilde S$. Analogous discussion holds also in the elliptical regime as well. We thus conclude that while the newly opened magnon channel can enhance the spin-mixing conductance (as shown by the positive corrections from $\mathcal{R}$ and $\mathcal{R}^c$), the suppression factor in front arising from the reduced order parameter may lead to an {\it overall} suppression in the interfacial spin transfer.\\

{\em Planar spin accumulation}.|For a spin accumulation parallel to the ferromagnet's easy plane, say $\bmus\parallel\ey$, the spin-transfer torque aligns the ferromagnetic 
spins uniformly and statically in the $y$ direction, in the steady state. While no superfluid spin transport is expected in this case, magnons can still carry spin current. Here, we consider magnon transport in the incoherent elastic transport limit, such that magnon current is conserved inside the ferromagnet for each energy channel, and the phase-relaxation length is much less than $L$~\cite{dattaBOOK95}.
In the circular-magnon regime, $T_D\ll T\ll T_c$, spin current (along the quantization axis $y$) is, furthermore, conserved across each interface. Therefore, the (outward) magnon current at interface $r$ is given by $J^s_r=(g_r^{\uparrow\downarrow}/\tS\pi)\int_0^\infty d\ve\rho^c_m(\ve)\ve_r[n_m(\ve)-n_{\rm BE}(\ve_r)]$, where $\ve_L=\ve-\mu_s$ and $\ve_R=\ve$. The spin-current injection at the left interface induces a nonequilibrium, energy-dependent magnon accumulation 
inside the ferromagnet. Imposing magnon current conservation at each energy in the ferromagnet, the nonequilibrium magnon distribution becomes:
\beq
n_m(\ve)=\frac{\ve_L g_L^{\uparrow\downarrow}n_{\rm BE}(\ve_L)+\ve_R g_R^{\uparrow\downarrow}n_{\rm BE}(\ve_R)}{\ve_L g_L^{\uparrow\downarrow}+\ve_Rg_R^{\uparrow\downarrow}}\,.
\eeq
Within linear response, we then obtain $J^s_R=(\mathcal{R}^c\mu_s/2\pi)g_L^{\uparrow\downarrow}g_R^{\uparrow\downarrow}/(g_L^{\uparrow\downarrow}+g_R^{\uparrow\downarrow})$.
Therefore, we find that the magnon current in this setup
is suppressed as temperature is lowered according to $J^s_R\propto {\mathcal{R}}^c\sim(T/T_c)^{3/2}$. A similar analysis for elliptical magnons at $T\ll T_D$ gives $J^s_R\propto(T/T_D)^{3/2}(T/T_c)^{3/2}$. It is interesting to note that the expression for the (outward) spin current into the metal reservoir at interface $r$, $J^s_r=(g_r^{\uparrow\downarrow}/\tS\pi)\int_0^\infty d\ve\rho^e_m(\ve)\ve_r[n_m(\ve)-n_{\rm BE}(\ve_r)]$, remains unmodified, using the appropriate magnon density of states, $\rho^e_m(\ve)=\ve^2/2\p^2 (\hbar v)^3$, for the elliptical magnons.

\end{document}